\documentstyle[twoside,fleqn,espcrc2,epsf]{article}

\newcommand{\AmS}{{\protect\the\textfont2
  A\kern-.1667em\lower.5ex\hbox{M}\kern-.125emS}}
\newcommand{\fb}{$f_B$}
\newcommand{\fbs}{$f_{B_s}$}

\newcommand{\bsixf}{$\beta=6.52$}

\def\3he{{$^3${\rm He}}}

\def\eg{{\it e.g.,\ }}

\def\slD{\raise.15ex\hbox{$/$}\kern-.53em\hbox{$D$}}
\def\dsl{\raise.15ex\hbox{$/$}\kern-.57em\hbox{$\Delta$}}
\def\slp{{\raise.15ex\hbox{$/$}\kern-.57em\hbox{$\partial$}}}
\def\nsl{\raise.15ex\hbox{$/$}\kern-.57em\hbox{$\nabla$}}
\def\sla{\raise.15ex\hbox{$/$}\kern-.57em\hbox{$\rightarrow$}}
\def\slla{\raise.15ex\hbox{$/$}\kern-.57em\hbox{$\lambda$}}
\def\slb{\raise.15ex\hbox{$/$}\kern-.57em\hbox{$b$}}
\def\lnp{\raise.15ex\hbox{$/$}\kern-.57em\hbox{$p$}}
\def\lnk{\raise.15ex\hbox{$/$}\kern-.57em\hbox{$k$}}
\def\lnK{\raise.15ex\hbox{$/$}\kern-.57em\hbox{$K$}}
\def\lnq{\raise.15ex\hbox{$/$}\kern-.57em\hbox{$q$}}

\def\cO{{\cal O}}

\def\pmb#1{\setbox0=\hbox{$#1$}%
\kern-.025em\copy0\kern-\wd0
\kern.05em\copy0\kern-\wd0
\kern-.025em\raise.0433em\box0 }

\def\q2{{Q^2}}
\def\gtwid{\raise.3ex\hbox{$>$\kern-.75em\lower1ex\hbox{$\sim$}}}
\def\ltwid{\raise.3ex\hbox{$<$\kern-.75em\lower1ex\hbox{$\sim$}}}
\def\12{{1\over2}}
\def\part{\partial}

\def\low#1{\lower.5ex\hbox{${}_#1$}}

\def\psl{\raise.15ex\hbox{$/$}\kern-.57em\hbox{$\partial$}}
\def\partt{\raise.15ex\hbox{$\widetilde$}{\kern-.37em\hbox{$\partial$}}}

\def\topppageno1{\global\footline={\hfil}\global\headline
={\ifnum\pageno<\firstpageno{\hfil}\else{\hss\twelverm --\ \folio
\ --\hss}\fi}}
 
\def\toppageno2{\global\footline={\hfil}\global\headline
={\ifnum\pageno<\firstpageno{\hfil}\else{\rightline{\hfill\hfill
\twelverm \ \folio
\ \hss}}\fi}}

\def\prd#1{Phys.\ Rev.\ {\bf D#1}}
\def\plb#1{Phys.\ Lett.\ {\bf #1B}}

\def\eg{{\it e.g.},\ }

\def\nsection#1 #2{\leftline{\rlap{#1}\indent\relax #2}}

\def\prd#1{Phys.\ Rev.\ {\bf D#1}}
\def\plb#1{Phys.\ Lett.\ {\bf #1B}}

\def\dallas{Nucl.\ Phys.\ {\bf B} (Proc.\ Suppl.) {\bf 34} (1994)}
\def\bielefeld{Nucl.\ Phys.\ {\bf B} (Proc.\ Suppl.) {\bf 42} (1995)}
\def\melbourne{Nucl.\ Phys.\ {\bf B} (Proc.\ Suppl.) {\bf 47} (1996)}
\def\stlouis{these proceedings}

\title{Update on $f_B$}

\author{ C.~Bernard,\hskip-0.03in
\address{{\vskip-0.10in{\hskip 0.07in Department of Physics, Washington
University, St.~Louis, MO 63130, USA}}} 
\thanks{presented by C.~Bernard}
T.~Blum,\hskip-0.03in
\address{Department of Physics, Brookhaven National Lab, Upton, NY 11973, USA}
T.~DeGrand,\hskip-0.03in
\address{Physics Department, University of Colorado, Boulder, CO 80309, USA} %
C.~DeTar,\hskip-0.03in
\address{Physics Department, University of Utah, Salt Lake City, UT 84112, USA}
Steven~Gottlieb,\hskip-0.03in
\address{Department of Physics, Indiana University, Bloomington, IN 47405, USA}
U.~M.~Heller,\hskip-0.03in
\address{SCRI, Florida State University, Tallahassee, FL 32306-4052, USA} 
J.~Hetrick,\hskip-0.03in
\address{Department of Physics, University of Arizona, Tucson, AZ 85721, USA} %
C.~McNeile,\hskip-0.03in$\,\null^{\rm d}$
K.~Rummukainen,\hskip-0.03in$\,\null^{\rm e}$
A.~Soni,\hskip-0.03in$\,\null^{\rm b}$
R.~Sugar,\hskip-0.03in
\address{Department of Physics, University of California, Santa Barbara, CA
93106, USA}
D.~Toussaint$\,\null^{\rm g}$
and M.~Wingate$\,\null^{\rm c}$
} 

\begin{document}

\begin{abstract}

We describe the current status of the MILC collaboration
computation of $f_B$, $f_{B_s}$, $f_D$, $f_{D_s}$ and their ratios.
Progress over the past year includes: better statistics and plateaus
at \bsixf\ (quenched), $\beta=5.6$ ($N_F=2$)
and  $\beta=5.445$ ($N_F=2$),
new runs with a wide range of dynamical quark masses at $\beta=5.5$ ($N_F=2$),
an estimate of the systematic errors due to the chiral extrapolation, and
an improved analysis which consistently takes into account both the
correlations in the data at every stage and the systematic effects
due to changing fitting ranges.
\end{abstract}

\maketitle


Existing and planned experimental measurements of
$B$-$\bar B$ and $B_s$-$\bar B_s$ mixing do not
constrain the Cabibbo-Kobayashi-Maskawa matrix without
knowledge of the heavy-light decay constants
\fb\ and \fbs\ and the corresponding $B$-parameters.
This fact has led to a major effort in the lattice
community to calculate these quantities \cite{cblat93}.

We have been computing
heavy-light decay constants with Wilson fermions
over the past three years. The goal of the current
stage of the computation is twofold: 1) to extrapolate the
quenched results to the continuum and estimate all systematic
errors within the quenched approximation, and 2) to estimate
errors due to quenching by comparing, at fixed lattice
spacing, results from quenched and from $N_F=2$ dynamical
lattices.

The key ingredients in the calculation have been described in \cite{lat94}.
Table~\ref{tab:lattices} gives the lattice parameters.
Computations on the largest lattices have been
performed on the 512-node and 1024-node Intel Paragon computers
at Oak Ridge National
Laboratory; Paragons at Indiana University and
at the San Diego Supercomputer Center have been used for smaller lattices.

We use the hopping parameter expansion of Henty and Kenway \cite{henty}
for the heavy quarks.
The static-light decay constants then come ``for free'' from the
computation.
However, as explained in \cite{lat94}, the procedure is optimized for
the heavy-light case, and the static-light results are not
usable when the physical volume is too large
(runs B, E, and L--P). A dedicated computation of the static-light
decay constants for these lattices is now in progress \cite{mcneile}.
We use a multi-source technique, with relative wavefunctions  taken from
the results of the Kentucky group \cite{wavefunctions}.
Since none of the  weak-coupling quenched lattices (and
only a subset of the coarser ones) are affected, the new running is unlikely to
have much impact on the extrapolation of the quenched results to the continuum.
However, the $N_F=2$ results at stronger coupling may be significantly altered.
At present, we thus base our estimate of quenching effects
on a comparison with the results
from run G, where the current static-light technique should be
reliable and indeed gives decay constants that are consistent
with the preliminary results from \cite{mcneile}.

\begin{table}
\caption{Lattice parameters.  Runs F, G, and L--P use
variable-mass Wilson valence quarks and
two flavors of fixed-mass staggered
dynamical fermions;
all other runs use quenched
Wilson quarks.}
\label{tab:lattices}
\centering
\begin{tabular}{cccc} \hline
name& $\beta\  (am_q)$ &size &\# configs. \\
\hline
\vrule height 10pt width 0pt A  & 5.7 & $8^3\times 48$ & 200 \\
\hline
\vrule height 10pt width 0pt B  & 5.7 & $16^3\times 48$ & 100 \\
\hline
\vrule height 10pt width 0pt E& $ 5.85$&  $12^3 \times 48$&  100 \cr
\hline
\vrule height 10pt width 0pt C  & 6.0 & $16^3\times 48$ & 100\\
\hline
\vrule height 10pt width 0pt D  & 6.3 & $24^3\times 80$ & 100\\
\hline
\vrule height 10pt width 0pt H& $ 6.52$&  $ 32^3 \times 100$& 60 \cr
\hline
\vrule height 10pt width 0pt F& $ 5.7\ (0.01)$&   $16^3 \times 32$&    49 \cr
\hline
\vrule height 10pt width 0pt G& $ 5.6\ (0.01)$&   $16^3 \times 32$&    200 \cr
\hline
\vrule height 10pt width 0pt L& $ 5.445\ (0.025)$&  $ 16^3 \times 48$&    100
\cr
\hline
\vrule height 10pt width 0pt N& $ 5.5\ (0.1)$&  $ 24^3 \times 64$&    100 \cr
\hline
\vrule height 10pt width 0pt O& $ 5.5\ (0.05)$&  $ 24^3 \times 64$&    100 \cr
\hline
\vrule height 10pt width 0pt M& $ 5.5\ (0.025)$&  $ 20^3 \times 64$&    100 \cr
\hline
\vrule height 10pt width 0pt P& $ 5.5\ (0.0125)$&  $ 20^3 \times 64$&    100
\cr
\hline

\hline
\vspace{-36pt}
\end{tabular}
\end{table}

A major improvement in the computation over the past year is the
consistent use of covariant fits at every stage of the data analysis.
Previously, only the fits of $f_{Qq} \sqrt{M_{Qq}}$ {\it vs.}\
$1/M_{Qq}$ took
correlations in the data into account.  As is well known \cite{kilcup},
the presence of
small, poorly determined eigenvalues in the covariance matrix
can make
covariant fits very unstable.
We therefore had been forced to use noncovariant fits to
the raw correlation
functions and in the chiral extrapolations.

Our new technique has many similarities to those proposed in \cite{kilcup},
but has some advantages, at least in the current project.
It is based on a standard approach in factor analysis \cite{factor}.
We first compute the correlation matrix (the covariance matrix normalized
with $1$'s along the diagonal) and find its eigenvalues and eigenvectors.
We then reconstruct the correlation matrix from the eigenvectors, but omitting
those corresponding to ``small'' eigenvalues.  The resulting
matrix is of course singular.  It is made into an acceptable correlation
matrix by restoring the $1$'s along the diagonal. Finally, the corresponding
covariance matrix is constructed and  inverted, and is used in the standard
way for making the fits.

The above technique interpolates smoothly between ordinary covariant fits,
where no eigenvalues are omitted, and noncovariant fits, where all
eigenvalues are omitted.  Furthermore, because the correlation
(as opposed to covariance) matrix
is used, the eigenvalues are normalized, with the average eigenvalue
always equal to 1.  This allows us to make a uniform determination
of which eigenvalues to keep, which is very important since we are dealing
with thousands of fits, and it is impossible to examine each fit by
hand.  Our standard procedure is to keep all eigenvalues greater
than 1; this accounts for typically $90$--$95\%$ of the total covariance.
Indeed, when one changes how the covariance matrix is computed,
(for example, by changing the number of configurations eliminated
in the jackknife), the eigenvalues smaller than 1 generally change
drastically.
The approach eliminates unstable, ``pathological'' fits completely.
We check that the final results are not significantly affected
when one keeps several more (or several fewer) eigenvalues throughout.

The use of covariant fits, with a reliable $\chi^2$ statistic,  has
led to three other important changes in the analysis over the past year.
First, we have found that the ranges in Euclidean time over which the
smeared-local and especially the smeared-smeared correlation functions
were previously fit were rather optimistic, giving $\chi^2/dof \sim 1.5$ with
40 or 50 degrees of freedom.
We now go out further in time to get reasonable confidence levels; this has
increased the statistical errors somewhat and
also  has tended to raise the central
values by about 1 (old) sigma.

\begin{figure}[htb]
\vspace{-105pt}
\epsfxsize=1.2 \hsize
\epsffile{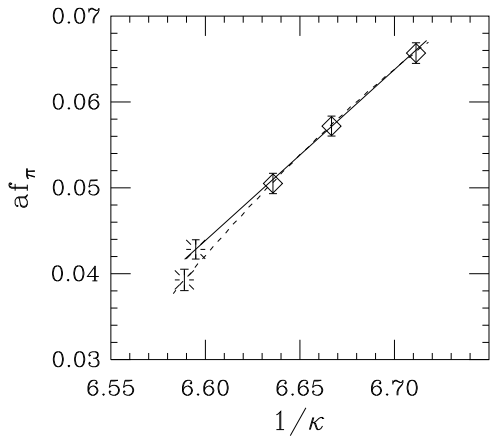}
\vspace{-28pt}
\caption{ Fits to $f_\pi$ {\it vs}.\  $1/\kappa$ for run D.
The linear fit
(solid line) has confidence level 0.03;
the quadratic ``fit'' (dashed line)
has no degrees of freedom.
The ``bursts'' show the extrapolation to $\kappa_c$, which in turn
is determined by the corresponding fit (linear or quadratic) to $m_\pi^2$
{\it vs}.\ $1/\kappa$.
}
\vspace{-18pt}
\label{fig:fpi_chiral}
\end{figure}

A second change involves the chiral extrapolations of $m_\pi^2$ and
$f_\pi$.  Fig.~\ref{fig:fpi_chiral} shows a typical chiral extrapolation
of $f_\pi$.  The standard linear fit in $1/\kappa$, although apparently
reasonable, has a poor confidence
level once correlations in the data
have been taken into account.  This feature also appears in fits
of $m_\pi^2$ {\it vs.}\ $1/\kappa$ and has been noted previously in the
literature \cite{rajan}.  Although there are no chiral logs in
$f_\pi$ in quenched chiral perturbation theory
at one loop \cite{cbmg},
nonlinear terms can enter at higher order.  However, the nonlinear terms
in Fig.~\ref{fig:fpi_chiral} could just as easily come from $\cO(a)$ effects.
For example, one can generally get acceptable confidence levels for
linear fits to $f_\pi$ simply by
using a different quark mass definition (\eg
$m_1(\kappa)\equiv \ln(1+ 1/2\kappa-1/2\kappa_c)$) on the
horizontal axis.
(Unfortunately, the curvature in $m_\pi^2$ increases with this definition.)
With just three light quark masses, we are unable to investigate
such higher order effects in detail.  We therefore choose linear fits
in finding the central values of decay constants, and take the
difference with the quadratic fits as a systematic error  estimate.
This is a significant ($\sim8\%$) error which we did not include
previously.  We note in passing that
linear fits for heavy-light
masses and decay constants as a function of light quark mass are generally
acceptable.

\begin{figure}[htb]
\vspace{-95pt}
\epsfxsize=1.2 \hsize
\epsffile{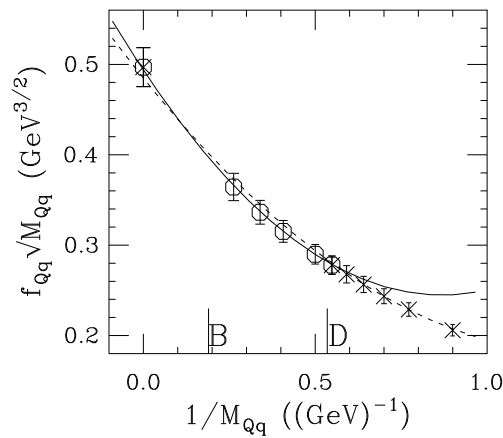}
\vspace{-28pt}
\caption{ $f_{Qq} (M_{Qq})^{1\over2}$ {\it vs}. $1/M_{Qq}$
for lattice D.  The solid line is a quadratic fit (conf.\ level = 0.87)
to the octagons  (``heavier heavies'' + static);
the dashed line is a quadratic fit (conf.\ level = 0.37)
to the crosses (``lighter heavies'' + static).
}
\vspace{-18pt}
\label{fig:frootm}
\end{figure}

A third change
appears in the interpolation
of $f_{Qq}\sqrt{M_{Qq}}$ in inverse meson mass to $m_B$.
A typical plot
is shown in Fig.~\ref{fig:frootm}.
 When all the fits at earlier stages
of the analysis are performed covariantly, the
fits to
the ``heavier heavies'' plus static generally have higher
confidence levels than fits to
the ``lighter heavies'' plus static and are therefore used for
the central values for quantities involving the $b$ quark.
On the other hand, when simple noncovariant fits
are used in the earlier analysis stages,
the two fits in Fig.~\ref{fig:frootm} typically have comparable
(and somewhat lower) confidence levels.  This was the case last year,
and we then chose the ``lighter heavies'' fit for the central values.
The effect on $f_B$ of the new choice varies with $\beta$
in the range $-3$ to $+10$ MeV.

\begin{figure}[htb]
\vspace{-85pt}
\epsfxsize=1.2 \hsize
\epsffile{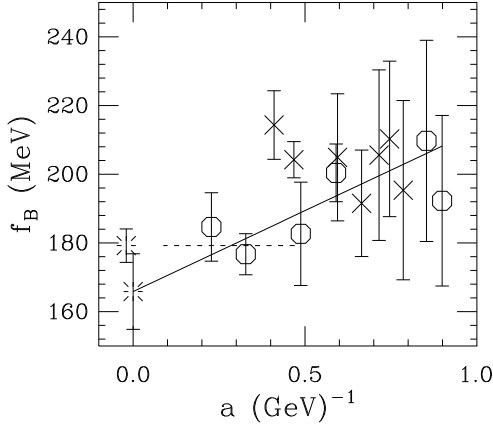}
\vspace{-28pt}
\caption{ $f_B $ {\it vs}.\ $a$.  Octagons are quenched data;
crosses, $N_F=2$.  The solid line is a linear fit to all
quenched points (conf.\ level$= 0.66$);  the dashed line is a a constant
fit to the three quenched points with $a<0.5$ GeV (conf.\ level$= 0.76$).
The extrapolated values at $a=0$ are indicated by
bursts.
The scale
is set by $f_\pi=132$ MeV throughout.
}
\vspace{-17pt}
\label{fig:fb}
\end{figure}

The extrapolation of quenched $f_B$  to the continuum is shown in
Fig.~\ref{fig:fb}. Errors on the points are larger than before \cite{lat94},
mainly because they now include, in addition to statistical errors,
the effects of varying the fitting ranges of the raw correlators.
We take the linear fit to all the quenched
data (solid line) to give the central value;  the dashed line
gives an error estimate.
An extrapolation of the $N_F=2$ data is not yet feasible, but may become
so when we complete 1) the new static computation on the coarser lattices
(L--P) and 2) new dynamical fermion runs at $\beta=5.6$, $24^3\times64$.

\begin{figure}[htb]
\vspace{-85pt}
\epsfxsize=1.2 \hsize
\epsffile{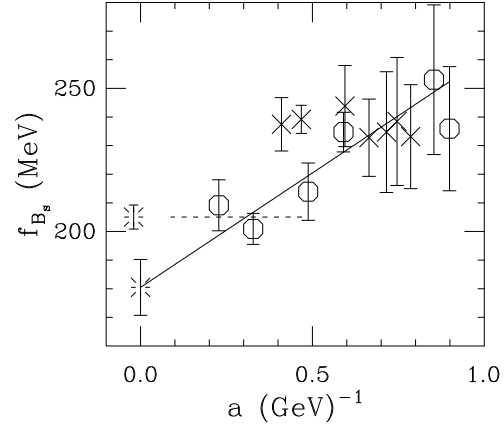}
\vspace{-28pt}
\caption{ $f_{B_s} $ {\it vs}.\ $a$.  Points
and fits as in Fig.~\ref{fig:fb}.
The solid line has
conf.\ level$= 0.37$;  the dashed line,
conf.\ level$= 0.48$.
}
\vspace{-17pt}
\label{fig:fbs}
\end{figure}

Figure~\ref{fig:fbs} is  a similar plot for \fbs.
Note that, although the difference at $a\sim0.5$ GeV between
quenched and $N_F=2$ results
for $f_B$ is not so clear, the corresponding difference seems
reasonably convincing for \fbs.  We are hopeful that the improvements
mentioned in the previous paragraph will sharpen these differences (if indeed
they are present).

The analysis of systematic errors is largely unchanged from {\it Lattice 95}
\cite{lat94}.
As described above, however, the errors due to chiral extrapolation are
now estimated and included.

The (still preliminary) results are:
\begin{eqnarray*}
& f_B  =  166(11)(28)(14)  \ & f_D  =   196(9)(14)(8) \\
& f_{B_s}  =  181(10)(36)(18) \   & f_{D_s}  =   211(7)(25)(11) \\
& {f_{B_s}\over f_B}  =   1.10(2)(5)(8) \ & {f_{D_s}\over f_D}  =
1.09(2)(5)(5)
\end{eqnarray*}
where the first error includes statistical
errors and systematic effects of changing fitting ranges; the second, other
errors
within the quenched approximation; the third, an estimate
of the quenching error. Decay constants are in MeV.
The estimate of the quenching error is crude at present and
may well increase significantly when a continuum extrapolation
of the dynamical fermion results is possible.

Computing was done at
ORNL Center for Computational Sciences, Indiana University, and SDSC.
We thank the Columbia group for supplying configurations F
and the HEMCGC collaboration for configurations G.
This work was supported in part by the DOE and NSF.

\end{document}